\documentclass[rnote]{aa}
\usepackage{txfonts}
\usepackage{graphicx}
\begin{document}
\title
% VORHER: Tautenburg NASPEC spectroscopy of 29 nearby star candidates
{Spectroscopic distances of 28 nearby star candidates\thanks{Based on observations 
with the 2m telescope of the Th\"uringer Landessternwarte Tautenburg and
with the 2.2m telescope of
the German-Spanish Astronomical Center, Calar Alto, jointly operated by 
the Max-Planck-Institut f\"ur Astronomie Heidelberg and the  
Instituto de Astrof\'{\i}sica de Andaluc\'{\i}a (CSIC).}}
% KORREKTURENDE
% VORHER: H. Meusinger, H. Jahrei\ss, R.-D. Scholz,
\author{H. Jahrei\ss\inst{1}\and
        H. Meusinger\inst{2}\and
        R.-D. Scholz\inst{3}\and
				B. Stecklum\inst{2}
% KORREKTURENDE
}

\offprints{H. Jahrei\ss}
\mail{jahreiss@ari.uni-heidelberg.de}
\institute{ Astronomisches Rechen-Institut am Zentrum f\"ur Astronomie der Universit\"at
Heidelberg, M\"onchhofstra{\ss}e 12-14, 69120
Heidelberg, Germany \and
Th\"uringer Landesssternwarte Tautenburg, Sternwarte 5, 07778
Tautenburg, Germany, \and
Astrophysikalisches Institut Potsdam, An der Sternwarte 16, 14482 Potsdam, Germany
}

\date{Printed: \today}

\abstract
{} 
% VORHER: We determined spectral types of 29 hitherto neglected nearby star candidates to improve our knowledge about their real properties.
{ 28 hitherto neglected candidates for the Catalogue of Nearby Stars (CNS)
were investigated to
verify their classification and to improve their distance estimates. All
targets had at least a preliminary status of being nearby dwarf stars based 
on their large proper motions and relatively faint magnitudes. Better 
photometric and/or spectroscopic distances were required for selecting 
stars which are worth the effort of trigonometric parallax measurements.} 
% KORREKTURENDE
% VORHER: Spectra were obtained with NASPEC at the Tautenburg 2m telescope. Spectra were reduced and spectral types were determined in comparison with stars of known spectral type.
{ Low-resolution spectra were obtained with NASPEC at the Tautenburg 2m 
telescope and with CAFOS at the Calar Alto 2.2m telescope. The spectral types 
of M-type stars were
determined by direct comparison of the target's spectra with those
of comparison stars of known spectral types observed with the same instrument.
The classification of earlier types was done based on comparison with
published spectral libraries.}
% KORREKTURENDE
{ For most of the target stars reliable spectral types could be determined 
and in combination with 2MASS photometry new improved distance estimates 
became available.
% NEU:
The majority were classified as M dwarfs including 11 stars within 
25\,pc. The fainter component of LDS\,1365, previously 
thought to form a nearby common proper motion pair, is according to our results 
an unrelated high-velocity background star. For several other nearby common 
proper motion pairs our distance estimates of the fainter components are in 
good agreement with Hipparcos distances of the brighter 
components.
The three stars in our sample which were previously thought to be white dwarfs
(GJ\,2091, GJ\,2094, GJ\,2098)
turned out to be in fact more distant high-velocity F- to K-type (sub)dwarfs.
For the star with the largest tangential velocity 
(GJ\,2091; $v_{\rm \,tan}>500$\,km/s) 
we have additional evidence for its probable 
Galactic halo membership from a measured large radial velocity 
of $266\pm25$\,km/s and 
from its $UBV$ photometry indicating a low metallicity.}
% ENDENEU
{}
\keywords{Galaxy: solar neighbourhood -- 
-- Galaxy: kinematics and dynamics} 

\titlerunning{Spectroscopic distances of 28 nearby star candidates}

\authorrunning{H. Jahrei\ss\ et al.}
\maketitle
%%%%%%%%%%%%%%%%%%%%%%%%%%%%%%%%%%%%%%%%%%

\section{Introduction\label{introd}}

In our efforts to complete our knowledge on nearby stars another sample 
of nearby star candidates could be successfully
observed at Tautenburg observatory 
% VORHER: in spring 2006
in the time span between spring 2006 and 2007.
% KORREKTURENDE
The candidate list contains without exception high-proper motion stars
that were as far as we know hitherto not yet observed spectroscopically. 
Spectra for these stars were desired not only to
complete the CNS (Catalogue of Nearby Stars) data sets but also to obtain first or improved distance estimates.

Most of the candidates were  in 
% VORHER: our CNS  data base 
the CNS data base managed by one of us (H.J.)
% KORREKTURENDE
since long ago. For example, the 
% VORHER: four
three
% KORREKTURENDE
white dwarf candidates originating from Eggen (\cite{ege68})
were already published in the GJ supplement (Gliese and Jahrei{\ss} \cite{gli79}). Presumably without further 
notice by
observers, at least no  information about the physical properties of these objects could be found in the literature since then.  

 Furthermore, some faint 
% VORHER: companions 
components
% KORREKTURENDE
of wide common
proper motion binaries were included in the observing list. Rather often the distance information for such stars is taken
from the brighter component. Yet, it is desirable to obtain independent distance information for the fainter 
component.

The remaining late type dwarfs, which were selected practically all from LHS, 
were included in the CNS data base only
after publication of the preliminary version of the 
Third Catalogue of Nearby Stars (pCNS3; Gliese and Jahrei{\ss} \cite{gli91}).
They were mostly border line
cases near or just beyond the present search limit of 25\,pc for the CNS sample. 

%%%%%%%%%%%%%%%%%%%%%%%%%%%%%%%%%%%%%%%%%%% 

\section{Observations and reduction \label{obs}}

% NEU:
Optical spectroscopic data were collected with the 
Nasmyth Focal Reducer Spectrograph (NASPEC) at the 2\,m telescope of 
the Th\"uringer Landessternwarte (TLS) Tautenburg. 
The V\,200 grism was used which yields a dispersion of
3.4\,{\AA} per pixel, or 12\,{\AA} FWHM for a slit width of 1\,arcsec.
The wavelength coverage was from about 4\,000\,{\AA} to  8\,500\,{\AA}.
For one star (GJ\,2091), 
the Tautenburg spectrum, taken under poor weather conditions, 
had not enough signal-to-noise for a proper classification. 
Therefore, we applied for director's discretionary time with 
the 2.2m telescope at Calar Alto. The Calar Alto focal reducer and 
faint object spectrograph CAFOS was used with the grism B\,200 giving
a wavelength coverage from about 3\,500\,{\AA}
to 7\,000\,{\AA} and a dispersion of 4.7\,{\AA} per pixel.
All spectra were reduced with standard routines from the ESO MIDAS data
reduction package. For details of the data reduction we refer to our
previous paper (Scholz et al. (\cite{sch05}).
% ENDENEU

The list of 
% VORHER: candidates observed in Tautenburg during February 2006 
nearby star candidates observed in Tautenburg during 
four runs between February 2006 and February 2007
% KORREKTURENDE
is presented in Table 1, with an internal CNS designation in column 1, the 2MASS right
ascension and declination for equinox J2000.0 in col 2  and 3, 
respectively, and the epoch of the 2MASS observation in column 4.
The annual proper motion and its 
direction is given in columns 5 and 6. These proper motions were practically all 
new determinations making use
of the available new digitized sky surveys like DSS and 2MASS
%(HARTMUT, die genauen 2MASS-Positionen waeren m.E. am besten und sollten
%zusammen mit deren Epochen in Tabelle 1 angegeben werden, oder ?? -
%HARMUT und HELMUT: ich habe die 2MASS-Identifikation aller
%Objekte gemacht und bin dabei auf Probleme mit NN1047AB =
%G148-47/G148-48 ?? gestossen - s. meinen email-Begleittext
%und das file: help-coords.lis). 
All our objects are at northern declinations, therefore
epoch differences of up to 50 years and more became
available with the USNO\,A2 positions  
% VORHER: of Poss\,I
from the first Palomar Observatory Sky Survey (POSS-I) 
% KORREKTURENDE
leading to proper motion errors well below 10\,mas/yr. In column 7 the 
previously known information about spectral
characteristics is given, i.e. listed are mostly spectral classes from Luyten's proper motion 
surveys. In column 8 
visual or photographic
(followed by a P) magnitudes are listed. In column 9 and 10 the previously known parallax and its standard error
are given, and in column 11 the source code for this parallax.  
 An additional 
identifier given in the last column should allow easier access to the Simbad data base for further information. 
\begin{table*} %[h]
\caption {Nearby candidates with new proper motions and previously known spectral characteristics and parallaxes}
\label{table:1}
%\centering
\begin{tabular}{lrrrrrrrrrclrrcl}
\hline \hline%\tbsp

 cns & & RA& \multicolumn{2}{c}{J2000.0} & Dec & & \multicolumn{1}{c} {epoch} &
\multicolumn{1}{c}{$\mu$} & \multicolumn{1}{c} {$\Theta$} &  SpT &  mag  & 
\multicolumn{1}{c}{$\pi$} & s.e. &source$^*$&  other   \\
 & h&m&\multicolumn{1}{c} {s}&\degr & \arcmin & \multicolumn{1}{c}{\arcsec} &  &\arcsec/yr & \degr \hphantom{3} & &  & mas  & & \\
\hline %\tbsp
% WEG: GJ\,2044   & 5& 38& 27&+11& 50.1 &0.213 & 164.6 & D?   &14.14: & 115.  &40.  &a&G 97-56   \\
GJ\,1167B  &13& 09& 41.96 & +29& 01& 56.9& 2000.277  &0.292 & 234.9 &  m   &17.6 P &      &      & &LDS 1365B \\
GJ\,2091   &12& 04& 56.32 & +04& 19& 56.0& 2000.137 &0.199 & 221.9 & WD?   &15.03  &  83. \hphantom{3} &27.\hphantom{3}  &a&G13-6     \\
GJ\,2094   &12& 37& 52.87 & +37& 30& 31.5& 1999.247 &0.140 & 208.2 & WD?   &15.62  &  60.\hphantom{3}  &20.\hphantom{3}  &a&GD 266    \\
GJ\,2098   &13& 08& 28.67 & +40& 27& 10.7& 1999.274 &0.263 & 242.2 & WD?   &14.00  & 160.\hphantom{3}  &52.\hphantom{3}  &a&G 164-46  \\
NN  817    &09& 53& 33.53 & +50& 45& 03.8& 1999.860 &0.530 & 272.0 &  m   &12.02  &  39.8 & 3.3 &h&LHS 2205  \\
NN  953    &11& 22& 36.29 & +35& 46& 19.0& 1998.189 &0.185 &   7.3 &  m   &14.62  &  41.3 & 9.3 &p&LP 264-44 \\
NN  991    &11& 47& 52.45 & +17& 42& 39.9& 1998.025 &0.217 & 130.2 &  m   &14.73  &  41.2 & 9.1 &p&LP 433-54 \\
NN 1047A   &12& 21& 27.05 & +30& 38& 35.7& 2000.263 &0.331 & 216.9 &  m   &15.31: &  81.7 &31.3 &p&G148-47   \\
NN 1047B   &12& 21& 26.73 & +30& 38& 37.6& 2000.263 &0.331 & 216.9 &  m   &15.4 : &  81.7 &31.3 &p&G148-48   \\
NN 1073    &12& 38& 36.21 & +05& 21& 38.5& 2000.184 &0.542 & 266.8 &  k-m &15.2 : &  38.4 &10.8 &p&LHS 2593  \\
NN 1083    &12& 43& 47.59 & +00& 03& 20.4& 1999.063 &0.691 & 224.8 &  m   &14.67  &  33.0 & 7.6 &p&LHS 2616  \\
NN 1169    &13& 39& 15.27 & +15& 40& 59.5& 1998.077 &0.980 & 142.3 &  m   &14.16: &  34.4 & 5.0 &t&LHS 2774  \\
NN 1246A   &14& 28& 21.52 & +05& 19& 01.4& 2000.263 &0.376 & 260.6 &  g-k &12.54  &  70.4 &51.5 &p&G 65-54   \\
NN 1246B   &14& 28& 17.58 & +05& 18& 45.9& 2000.263 &0.378 & 259.3 &  g-k &13.13  &  70.4 &51.5 &p&G 65-53   \\
NN 1342B   &15& 35& 25.67 & +60& 05& 07.7& 1999.348 &0.231 & 132.9 &  m   &13.46  &  52.4 &  .7 &h&LP 99-392 \\
NN 1356    &15& 49& 36.74 & +00& 17& 48.9& 2000.326 &0.599 & 216.9 &  m   &13.20  &  35.9 &10.5 &p&LHS 3123  \\
NN 1361A   &15& 50& 18.62 & +34& 37& 11.5& 1998.255 &0.195 & 222.6 &  m   &13.04: &  37.1 & 8.3 &p&LP 274-11 \\
NN 1361B   &15& 50& 18.55 & +34& 37& 15.9& 1998.255 &0.195 & 222.6 &  m   &16.5 : &  37.1 & 8.3 &p&LP 274-10 \\
NN 1538B   &18& 00& 45.44 & +29& 33& 56.7& 1999.260 &0.222 & 321.2 &  k   &14.5 P &  34.6 &  .7 &h&LP 333-29 \\
nn   40    &08& 12& 26.57 & +43& 09& 29.0& 1998.852 &0.558 & 165.2 &  k-m &14.26  &  38.3 & 8.7 &p&LHS 1988  \\
nn   44    &09& 08& 50.79 & +54& 07& 18.3& 1999.948 &0.553 & 221.4 &  m   &15.04  &  42.3 & 9.1 &p&LHS 2109  \\
nn   51    &11& 00& 35.69 & +37& 28& 50.8& 1998.271 &0.724 & 251.8 &  k-m &15.36  &  42.9 & 9.1 &p&LHS 2340  \\
nn   76    &13& 32& 00.07 & +66& 36& 00.5& 1999.263 &0.567 & 300.5 &  m   &14.05  &  40.6 & 9.4 &p&LHS 2748  \\
nn   77    &14& 07& 48.03 & +57& 11& 45.3& 1999.123 &0.693 & 285.4 &  m   &14.20  &  47.1 &10.5 &p&LHS 2864  \\
nn   78    &14& 16& 42.11 & +36& 00& 50.8& 2000.186 &0.542 & 146.1 &  k-m &12.59  &  33.9 & 9.2 &p&LHS 2886  \\  
nn   79    &14& 20& 39.46 & +46& 02& 34.2& 1998.395 &0.645 & 253.6 &  m   &15.10  &  45.4 & 9.6 &p&LHS 2897  \\
nn   87    &15& 03& 35.01 & +84& 38& 26.7& 1999.296 &0.578 & 221.0 &  m   &14.15  &  33.7 & 7.6 &p&LHS 3017  \\
nn   90    &16& 34& 13.72 & +54& 23& 39.8& 1999.356 &0.569 & 337.5 &  k   &11.86  &  33.9 & 9.1 &p&LHS 3214  \\
\hline
\end{tabular}

$^*$  a: WD by Eggen(1968), p: photometric parallax, h:
Hipparcos (ESA \cite{esa97}) parallax, and t: ground-based trigonometric parallax.
%\end{center}
\end{table*}

%HARMUT, ist die Bezeichnung fuer vermutete weisse Zwerge 
%nicht ''WD?'' anstelle von ''D?'' (Tabelle 1) ??

% NEU:
\begin{figure}
  \resizebox{\hsize}{!}{\includegraphics{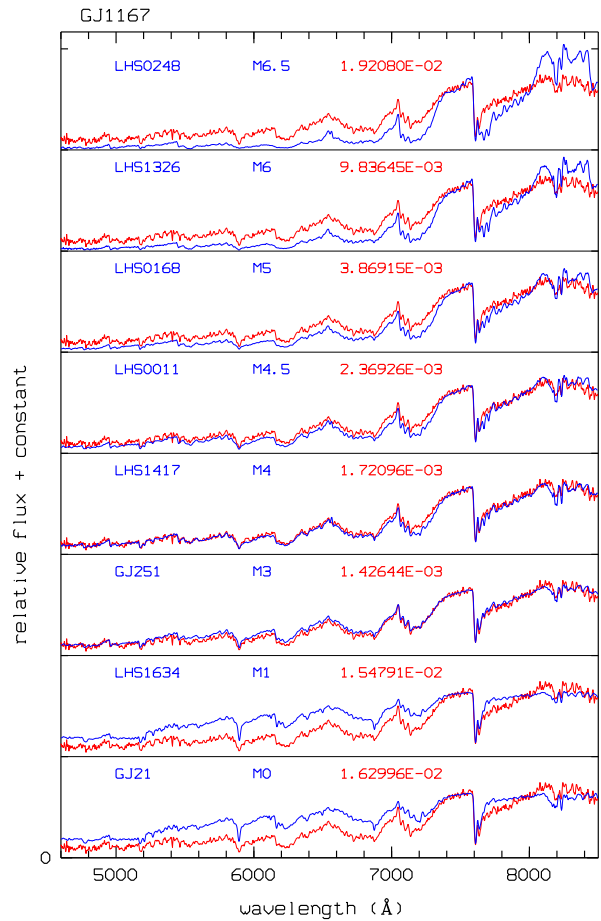}}
  \caption{Spectrum of GJ\,1167B in comparison to a sequence of M dwarfs
           with known spectral types. The numbers give the variance
           between the target spectrum and a given comparison spectrum.
           The signal-to-noise of the NASPEC spectra shown here are typical of
           all our M-type targets.}
  \label{fig_spdiff}
\end{figure}
% ENDENEU

Additional spectra of nine stars with known spectral types from M0 to M6.5 
% NEU:
(see Figure~\ref{fig_spdiff})
% ENDENEU
were taken to allow a direct comparison of our 
target spectra. 
Unfortunately, it turned out only later on that we could not make use of 
the spectrum taken for the M2 dwarf
 Gl\,38. Therefore a comparison spectrum for M2 dwarfs is missing in our sequence. For the M dwarf stars the
spectral type then was estimated by comparing the spectra in question directly with the sequence of comparison
spectra. Hereby all spectra were normalized at 7500\,{\AA}. 
% NEU:
Figure~\ref{fig_spdiff} shows as an example the spectrum of GJ\,1167B
compared to the spectra of eight comparison stars.
% ENDENEU
The visual impression allows a rather accurate determination
of the spectral type. Yet, it must be taken into account that 
the calibration of the spectra below 5000\,{\AA} and above 
8000\,{\AA} is less certain.

In addition to the visual estimate 
% VORHER: a computerised automatic 
an automatic classification
% KORREKTURENDE
method was applied. The difference between the
spectrum of the object and every comparison spectrum was 
calculated and the variance was used as a measure of
agreement. The best fit spectral type 
% VORHER: then can be  obtained as minimum of the variance.
was then obtained by searching the minimum of the variance as a 
function of spectral type using a quadratic regression.
% KORREKTURENDE 
% NEU:
Generally, the automatic method provided results in good agreement with
the visual classification. Nevertheless, we preferred the results of the 
latter.
% ENDENEU
 
%\include {table1}

%%%%%%%%%%%%%%%%%%%%%%%%%%%%%%%%%%%%%%%%%%%

% VORHER: Results and discussion
\section{Object classification and distance estimates \label{res}}
% KORREKTURENDE

In Table 2 the finally adopted spectral types are given in column 2. 
% VORHER: These
For all the M dwarfs, these spectral types 
% KORREKTURENDE
were obtained by visual 
inspection comparing the
target spectra with the sequence of comparison spectra. 
Whereas column 3 lists spectral types obtained by the automatic method described above in section \ref{obs}. 
% NEU:
The adopted spectral types in column 2 are extended by the letter ''e''
if an H$\alpha$ emission line was detected. It is interesting that all
emission line M dwarfs were also identified as X-ray sources according
to the VizieR Catalogue Service (http://vizier.u-strasbg.fr/).
% ENDENEU
% VORHER: The 2MASS colours J, H, and K 
The 2MASS magnitude $J$ 
% KORREKTURENDE 
in column 4 was used to compute the spectroscopic parallaxes listed in column 5 with the recently calibrated spectral type vs. $M_J$
relation for late type dwarf stars given in Table 2 of Scholz et al. (\cite{sch05}). 
\begin{table} %[h]
\caption {New spectral types, 2MASS photometry, and resulting spectroscopic parallaxes }
%\centering
%\begin{tabular}{l@{\extracolsep\fill}ccrrrrl}
\begin{tabular}{lccrrl}
\hline \hline %\tbsp

   cns    & SpT    & SpT$_c$ &
\multicolumn{1}{c}{$J$} & $\pi_{sp}$ &  other   \\
\hline %\tbsp
GJ\,1167B& M3.5 & M3.4 &13.733& 6.5 &LDS 1365B \\
% WEG: GJ\,2044  & G?   &      &12.755&12.414&12.313&  &G 97-56   \\
GJ\,2091  & $<$(sd)G0 &      &13.667& $<$1.8 &G13-6     \\
GJ\,2094  & $\sim$(sd)K5 &      &14.406& $<$2.9 &GD 266    \\
GJ\,2098  & $<$(sd)K0 &      &12.405& $<$4.9 &G 164-46  \\
NN  817  & M1.5  & M1.5 & 8.733&  37.8 &LHS 2205  \\
NN  953  & M4.0  & M3.7 &10.243&  42.2 &LP 264-44 \\
NN  991  & M3.0  & M2.8 &10.406&  24.7 &LP 433-54 \\
NN 1047A & M4.5e & M3.9 & 9.987&  59.2 &G148-47   \\
NN 1047B & M5.0e & M6.1 &10.057&  70.6 &G148-48   \\
NN 1073  & M4.5  & M4.5 &10.768&  41.3 &LHS 2593  \\
NN 1083  & M4.0  & M3.2 &10.420&  38.9 &LHS 2616  \\
NN 1169  & M3.5  & M3.4 &10.333&  31.3 &LHS 2774  \\
NN 1246A & M3.5  & M3.3 & 8.724&  65.6 &G 65-54   \\
NN 1246B & M3.0  & M2.4 & 8.932&  48.7 &G 65-53   \\
NN 1342B & M4.0e & M3.4 & 9.270&  66.1 &LP 99-392 \\
NN 1356  & M3.0  & M2.6 & 9.464&  38.1 &LHS 3123  \\
NN 1361A & M3.0  & M2.8 & 9.414&  39.0 &LP 274-11 \\
NN 1361B & M4.5e & M4.4 &11.025&  36.7 &LP 274-10 \\
NN 1538B & M2.0  & M1.9 & 9.058&  36.8 &LP 333-29 \\
nn   40  & M3.5  & M2.6 & 9.994&  36.6 &LHS 1988  \\
nn   44  & M4.0  & M3.0 &10.530&  37.0 &LHS 2109  \\
nn   51  & M4.5  & M4.4 &10.659&  43.5 &LHS 2340  \\
nn   76  & M3.5  & M3.6 & 9.916&  37.9 &LHS 2748  \\
nn   77  & M4.0  & M3.2 & 9.875&  50.0 &LHS 2864  \\
nn   78  & M2.0  & M2.0 & 9.183&  34.8 &LHS 2886  \\
nn   79  & M4.5  & M4.3 &10.455&  47.8 &LHS 2897  \\
nn   87  & M4.0  & M3.2 &10.071&  45.7 &LHS 3017  \\
nn   90  & M0.0  & M1.1 & 8.720&  29.0 &LHS 3214  \\
\hline
\end{tabular}
%\end{center}
\label{results}
\end{table}
%\include {table2}

%HARTMUT, eigentlich verwenden wir die $H$ und $K_s$ aus dem 2MASS
%doch gar nicht und koennten sie deshalb in Tabelle 2 weglassen. Damit
%waere diese Tabelle wahrscheinlich auf eine Spalte begrenzt, und das
%Paper waere ev. auf 5 Seiten begrenzt. Was haeltst Du von dieser Kuerzung ??

% NEU:
\begin{figure}
  \resizebox{\hsize}{!}{\includegraphics[angle=270]{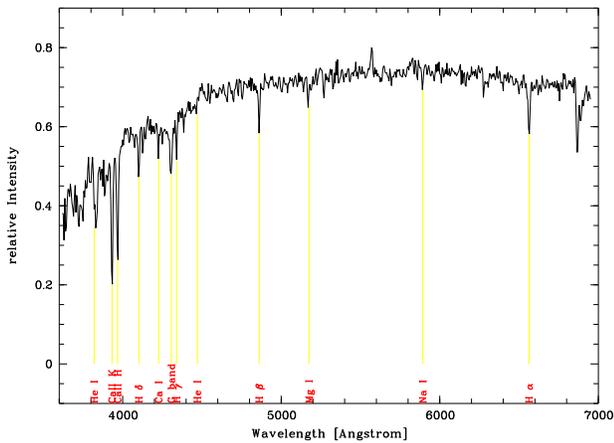}}
  \caption{CAFOS B200 spectrum of GJ\,2091.}
  \label{fig_gj2091}
\end{figure}
\begin{figure}
  \resizebox{\hsize}{!}{\includegraphics[angle=270]{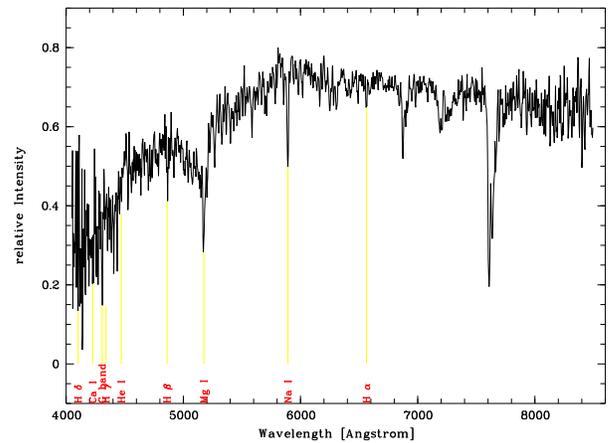}}
  \caption{NASPEC V200 spectrum of GJ\,2094.}
  \label{fig_gj2094}
\end{figure}
\begin{figure}
  \resizebox{\hsize}{!}{\includegraphics[angle=270]{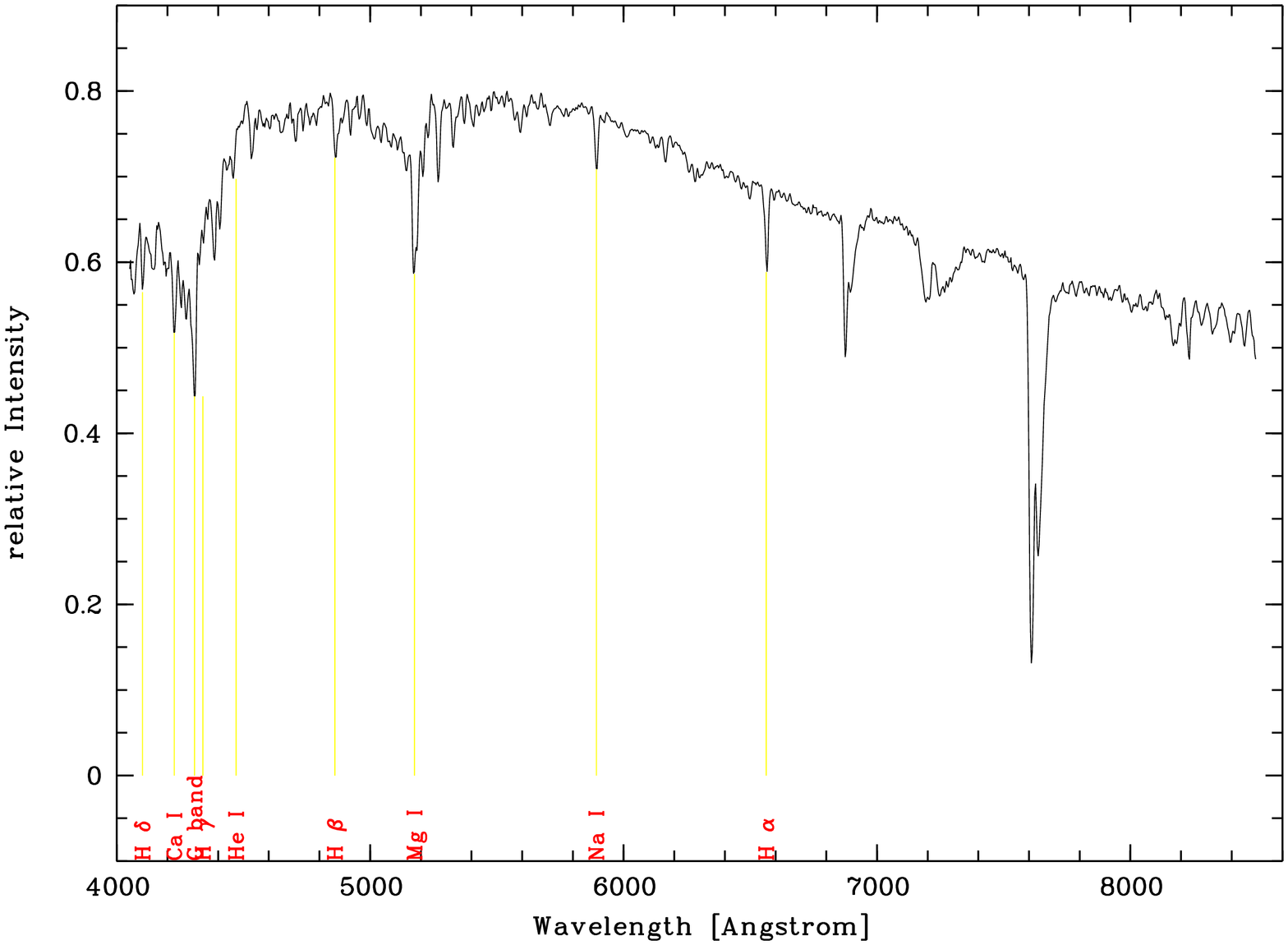}}
  \caption{NASPEC V200 spectrum of GJ\,2098}
  \label{fig_gj2098}
\end{figure}

The spectra of the three previously suspected white dwarfs GJ\,2091, GJ\,2094, and GJ\,2098 are shown in
Figure~\ref{fig_gj2091},~\ref{fig_gj2094},
and ~\ref{fig_gj2098}, respectively. These are clearly no white dwarf
spectra but rather typical for F- to K-type dwarfs 
%{\bf Hier war ich beim Lesen etwas verwirrt. Worauf bezieht sich das "earlier"?}
although in the case of 
GJ\,2091 we were initially not sure. The NASPEC spectrum of this object
(not shown here) was very noisy allowing only the identification
of the H$\alpha$ and H$\beta$ lines. But no other, even broad
stellar features could be seen. However, the higher-quality CAFOS spectrum 
(Figure~\ref{fig_gj2091}) ruled out a white dwarf nature. The 
signal-to-noise ratio of the NASPEC spectrum of GJ\,2094 (Figure~\ref{fig_gj2094})
is as low as in the poor NASPEC spectrum of GJ\,2091, but broad
absorption to the blue side of the MgI line (due to MgH) and a strong 
NaI line are clearly seen in addition to H$\alpha$ and H$\beta$.

For our objects of earlier spectral types ($<$M) we had no comparison objects
observed with the same instrument. In order to classify them
we searched for similar spectra in the spectral libraries of 
Torres-Dodgen \& Weaver (\cite{torres93}) and Jacoby et al. (\cite{jacoby84}).
Based on the strong NaI and weak H$\alpha$ we classify GJ\,2094 as
a mid-K dwarf. It could well be a subdwarf since the spectrum looks very
similar to that of the high-proper motion star SSSPM~J1549$-$3544 classified 
as sdK5 by Farihi et al. (\cite{farihi05}). 
In the case of GJ\,2098 the NaI line is much weaker compared
to H$\alpha$, which is typical for G dwarfs. The strong Balmer lines
together with rather weak MgI and NaI lines suggest an even earlier,
F-type classification for  GJ\,2091. 

If we assume all three stars GJ\,2094, GJ\,2098, and GJ\,2091 to be on the main sequence and adopt a
rough classification as $\sim$K5, $<$K0, and $<$G0, 
we obtain very large distances of 
680\,pc, $>$410\,pc, and $>$1100\,pc, respectively, by using again the absolute 
$J$ magnitudes for K5 and K0 stars from Scholz et al. (\cite{sch05})
and an $M_J=3.4$ for a G0 star (Boccaletti et al. \cite{boccaletti03}).
Consequently, these distances lead to extremely large tangential velocities of
450\,km/s, $>$500\,km/s, and $>$1000\,km/s, respectively. If all three
objects were subdwarfs, their distances and tangential velocities
would be reduced by roughly 50\% (Scholz et al. \cite{sch05}).
In any case, all three objects are far beyond the Solar neighbourhood
and possess very high velocities typical for the Galactic halo population.

% ENDENEU

%%%%%%%%%%%%%%%%%%%%%%%%%%%%%%%%%%%%%%%%%%%

\section{Individual objects \label{indiv}}

In the following some of the most interesting objects are discussed in more detail.

%%%%%%%%%%%%%%%%%%%%%%%%%%%%%%%%%%%%%%%%%%%

\subsection{GJ\,1167B \label{gj1167}}

In his second list of common proper motion pairs W. J. Luyten (\cite{luy69}) published LDS 1365
with a common proper motion of 0.39\arcsec/yr in 240\degr and a large separation of 190\arcsec in
26\degr. A spectral class "m" was assigned to both components. In a subsequent publication Luyten (\cite{luy73}) designated 
the brighter component with LP\,322-836  
and the fainter one with LP\,322-835. The couple was included in the GJ-supplement as GJ\,1167A and B on the basis of
 a photometric parallax of  54\,mas for the brighter star LP\,322-836. A first 
trigonometric parallax from Sproul Observatory 
(Heintz \cite{hei94}) 
$\pi = 86 \pm 12.5$\,mas moved it even closer to
the Sun. This value was recently confirmed by Smart et al. (\cite{sma07}) who determined the trigonometric parallax
of LP\,322-836 to $\pi = 83.1 \pm 6.2$\,mas.

Sanduleak's (\cite{san76}) remark (see star No. 224) that he found a much brighter non-M star
at Luyten's position for LP\,322-835 
 may have  caused Luyten to omit LP\,322-835 from his
NLTT-catalog, and so it was also excluded from the pCNS3. 

New proper motion determinations based on the modern DSS etc. yield $\mu  = 0.398\arcsec/yr$ in 239\fdg1 for 
LP\,322-836  and  $\mu  = 0.291\arcsec/yr$ in 234\fdg8 for LP\,322-835, and first doubts about
a physical connection appeared. Now our spectral type M3 for LP\,322-835 allowed a first direct distance determination 
for component B to
190\,pc. This is in close agreement with a distance of 175\,pc which we derived simultaneously from Sloan colours
i-z applying the calibration given in Hawley et al. (\cite{haw02}). In other words LDS 1365 
does not 
form a common
proper motion pair. LP\,322-836 --  henceforth GJ\,1167 -- is a nearby slow moving young M4 dwarf showing X-ray
emission. On the other hand the formerly supposed common proper motion companion LP\,322-835 is a far away M3 dwarf with
 a rather  high ($\sim 250$\,km/s) tangential velocity.

%%%%%%%%%%%%%%%%%%%%%%%%%%%%%%%%%%%%%%%%%%%

\subsection{GJ\,2091}

In an investigation of faint blue stars Eggen (\cite{ege68}) suspected this high-proper motion star to be a white dwarf. 
On the basis  of his $UBV$
photometry he deduced a parallax of 83\,mas in applying an $U-V, M_V$\,relation.  With this information G\,13-6
was included in Table 2 (Suspected Nearby Stars) of the GJ-supplement Gliese and Jahrei{\ss} (\cite{gli79}). 
Since then not any notice was taken of 
this object  in the literature. This applies also to the 
% VORHER: three stars  GJ\,2044, 2094, 2098, 
two stars GJ\,2094 and GJ\,2098, 
% KORREKTURENDE
with a similar history.

% WEG: Contrary to the latter GJ-stars which according to our new spectra turned out to be most probably G-type stars,
% VORHER: the spectrum of GJ\,2091 obtained at Tautenburg observatory under poor weather conditions let us a preliminary classification that it may be
The NASPEC spectrum of GJ\,2091 was obtained under poor weather conditions, and led 
us to a preliminary classification of this object as 
% KORREKTURENDE
a cool DA white dwarf with a clear indication of H${\alpha}$ and 
H${\beta}$ absorption lines and no further significant features. In this case GJ\,2091 would have a distance close 
to 10 parsec.

 Yet, for a definitive decision a better spectrum was required, and we applied for DDT service observation with
CAFOS at the 2.2m telescope at Calar Alto.
Such an observation was successfully carried out on April 23, 2007. 
The analysis of this spectrum (Figure~\ref{fig_gj2091}) shows clearly 
that GJ\,2091 is not a
white dwarf but rather a F-G dwarf star so that it can definitively be excluded from the CNS. Nevertheless we determined a radial
velocity for this object,
% NEU:
based on measurements of the four Balmer lines H$\alpha$,..., H$\delta$ with
respect to the spectrum of the standard star, Feige\,66, observed 
during the same night. The radial velocity
% ENDENEU
turned out to be with $+266 \pm 25$km/s rather large, i.e. GJ\,2091 
is really a fast moving high velocity star, 
% NEU:
probably a member of the Galactic halo.
The available $UBV$ photometry of GJ\,2091 points at a low metallicity
[Fe/H]\,=\,--2.4 according to Carney (\cite{carney79}), which is also
an indication of a membership in the Galactic halo.
% ENDENEU
%%HARTMUT, ich habe hier Daten aus Deiner email (Date: 
%%2007-05-10 15:10) verwendet, ok ??

%%%%%%%%%%%%%%%%%%%%%%%%%%%%%%%%%%%%%%%%%%%

\subsection{LHS\,2205 = NN 817}

With a Hipparcos parallax of $39.75 \pm 3.27$\,mas LHS\,2205 is just very close to the search limit of CNS. It was missed by former 
spectroscopic
 surveys due to its slightly smaller photometric parallax of 38\,mas, the reason that it was excluded from the 
pCNS3.
 Its new spectral type M1.5 yields a spectroscopic parallax of 38\,mas in full agreement with the photometric value and the
trigonometric parallax measured by Hipparcos.

%%%%%%%%%%%%%%%%%%%%%%%%%%%%%%%%%%%%%%%%%%%

\subsection{LP\,433-54 = NN 991}

  Optical photometry yields a distance of 24\,pc, whereas infrared photometry as well as the new M3.0 spectral type 
predict a considerably larger distance up to 
% VORHER: 33\,pc.
40\,pc.
% KORREKTURENDE

%%%%%%%%%%%%%%%%%%%%%%%%%%%%%%%%%%%%%%%%%%%

\subsection{G\,148-47/48 = NN 1047AB}

 A first distance of 14\,pc  for this close pair (separation 4.5\arcsec) 
 was determined by Pesch and Dahn (\cite{pes82})  calling it Sanduleak 57+58. The authors  
used for their estimate $VI$ photometry $V = 14.51, V-I = 3.40 $ where the $V-I$ was measured in the Kron-Mayall system.
Transforming this value to the Kron-Cousins system according to Leggett (\cite{leg92}) yields only a minor change to
  $ V-I = 3.39$.
 Fleming
(\cite{fle98}) published an independent distance estimate of 11.2\,pc from his slightly different $VI$ photometry
 (in the Kron-Cousins system) $ V = 14.709, V-I = 3.465$ without being aware that it is
a close pair. The latter was recently 
% VORHER: rediscussed in Reid et al. (\cite{rei04}), obtaining 
discussed by Reid et al. (\cite{rei04}), who obtained
% KORREKTURENDE
a
revised distance estimate for the system of $\sim 12.5$\,pc. 

% VORHER: Combining the spectroscopic distance estimates for the two components with the photometrically determined estimates we obtain a distance of 17\,pc under the assumption that the two components are of almost equal brightness. 

Averaging our independent spectroscopic distance estimates for the two 
components we obtain a mean distance of 17\,pc for the system. This is close 
to a corrected value of 15.8\,pc which one gets from the original estimate of
Fleming (\cite{fle98}) assuming a system 
of almost equal brightness components, which is supported by several observers.
%HARTMUT, ich habe Deinen Satz vorher nicht verstanden, Du hast doch nur 
%gemittelt, um zu den 17 pc zu kommen, oder ??
% KORREKTURENDE 

%%%%%%%%%%%%%%%%%%%%%%%%%%%%%%%%%%%%%%%%%%%

\subsection{G\,65-54/53 = NN 1246AB }

This common proper motion pair separated by 61 arcsec had up to now only 
$UBV$-photometry with $B-V > 1.50$ for both components.
Because a value of  $B-V \ge 1.30$  doesn't allow a reasonable 
distance estimate this couple was excluded from the pCNS3. Now with the new spectral types 
and 2MASS photometry a distance of 17.5\,pc seems to be indicated.  

%%%%%%%%%%%%%%%%%%%%%%%%%%%%%%%%%%%%%%%%%%%

\subsection{LP\,99-392 = NN 1342B }

LP\,99-392 is the fainter common proper motion companion (separation 43\arcsec) to
HD\,139\,477. Here we had the curious case that a $VRI$-photometric parallax of 55\,mas  was available for the fainter
component, whereas for the brighter component only the HD-type K5 was known. Therefore the pair wasn't included in the
pCNS3 though it was already then taken into consideration.

 Only in 1997 Hipparcos determined a distance of 19\,pc for HD\,139\,477, and
almost simultaneously a MK spectral type of K3\,V was published for it (Loth and Bidelman \cite{lot98}). The
combination of photometric and spectroscopic parallaxes yields a distance of 18\,pc for LP\,99-392, and also its proper
motion of  $\mu = 0.231$\arcsec/yr in 132\fdg9 is in good agreement with the Hipparcos value $\mu = 0.2370$\arcsec/yr in 133\fdg4 
for  HD\,139\,477. In other words the physical connection of this 
% VORHER: cpm 
common proper motion
% KORREKTURENDE
pair is now well established.

%%%%%%%%%%%%%%%%%%%%%%%%%%%%%%%%%%%%%%%%%%%

\subsection{LP\,333-29 = NN 1538B }

Already the newly determined proper motion $\mu = 0.222$\arcsec/yr in 321\fdg8 for LP\,333-29 allowed little doubts
about the physical connection to the G2\,V primary HD\,164595 about 88\arcsec away having a Hipparcos proper motion of 
$\mu = 0.2224$\arcsec/yr in 321\fdg2. Now our spectral type of M2.0 provides a
direct distance estimate for LP\,333-29 to 27.2\,pc, which is in almost perfect agreement with the Hipparcos parallax of
34.57\,mas for the primary, and the metallicity $[Fe/H] = -0.07$ determined for the primary in Bonfils et al.
(\cite{bon05}) applies also to the M2 dwarf LP\,333-39.

%%%%%%%%%%%%%%%%%%%%%%%%%%%%%%%%%%%%%%%%%%%

\subsection{LHS\,1988 = nn 44 }

$VRI$ photometry provides consistently a parallax of 42\,mas. 
Whereas $V-K_s$ with 38\,mas and the new spectral type M4.0
with 37\,mas yield slightly smaller values.

%%%%%%%%%%%%%%%%%%%%%%%%%%%%%%%%%%%%%%%%%%%

%\subsection{LHS\,1988 = nn 44 \label{lhs1988}}

%%%%%%%%%%%%%%%%%%%%%%%%%%%%%%%%%%%%%%%%%%%

\section{Conclusions \label{concl}}

With the 
% VORHER: spectra published in this paper, 
spectroscopic distances determined in this paper,
% KORREKTURENDE
we were able to solve several 
doubtful distance estimates for nearby star
candidates. Among these it turned out that Eggen's white dwarf 
suspects contained in the GJ-supplement are all far
away F-K dwarf stars. 
% NEU:
From their extremely large tangential velocities we suspect them to
be in fact subdwarfs. With this classification they would still have
large velocities typical for Galactic halo stars. Our measured radial
velocity for the star with the largest tangential velocity,
GJ\,2091, confirms this assumption.

All other stars in our sample were classified as M dwarfs. 
According to our new distance estimates
all but one of these 25 stars lie within 40\,pc, 
including 11 objects falling within
the 25\,pc horizon of the CNS.

% ENDENEU

We could further proof that Luyten's common proper motion pair LDS\,1365 consists in reality of two M dwarf stars with similar
proper motions but completely different distances and space velocities. 

On the other hand our spectroscopy in combination with 2MASS colours provided a first reliable distance estimate 
for the common proper motion pair G\,65-54/53, which is probably within 20\,pc.    
%%%%%%%%%%%%%%%%%%%%%%%%%%%%%%%%%%%%%%%%%%%

\section*{Acknowledgments}
 
% NEU:
We thank the Calar Alto Observatory for allocation of director's discretionary 
time to this programme.
We acknowledge the
use of the Simbad database, the VizieR Catalogue Service operated
at the  CDS, and the use of the 2MASS All-Sky Survey. 
% ENDENEU

%%%%%%%%%%%%%%%%%%%%%%%%%%%%%%%%%%%%%%%%%%%%%%%%%%%%%%%%%%%%%%%%%%%%%%%%

\end{document}